\documentstyle[12pt]{article}
\font\grande=cmr10 scaled \magstep4
\font\medio=cmr10 scaled \magstep2

\def\laq{\raise 0.4 ex \hbox{$<$}\kern -0.8 em\lower 0.62 ex\hbox{$\sim$}}
\def\gaq{\raise 0.4 ex \hbox{$>$}\kern -0.7 em\lower 0.62 ex\hbox{$\sim$}}

\def\laq{\raise 0.4ex\hbox{$<$}\kern -0.8em\lower 0.62
ex\hbox{$\sim$}}
\def\gaq{\raise 0.4ex\hbox{$>$}\kern -0.7em\lower 0.62
ex\hbox{$\sim$}}

\begin{document}
\bibliographystyle {unsrt}
\newcommand{\pa}{\partial}

\titlepage
\begin{flushright}
DF/IST--2.96 \\
\end{flushright}
\vspace{15mm}

\begin{center}
{\bf Topological  Inflation in Dual Superstring Models}\\
\vspace{5mm}
{\grande }
\vspace{10mm}
M. C. Bento and O. Bertolami\\
{\em Departamento de F\'{\i}sica,
  Instituto Superior T\'ecnico}\\ 

{\em Av. Rovisco Pais, 1096 Lisboa Codex, Portugal } \\
\end{center}

\vspace{10mm}
\centerline{\medio  Abstract}
\noindent
We study the possibility of obtaining inflationary solutions from S-dual 
superstring potentials. We find, in particular, that such solutions occur 
at the core of domain walls separating degenerate minima whose positions 
differ by modular transformations.\\

\vfill
\begin{flushleft}
DF/IST--2.96 \\
May 1996 \end{flushleft}

\newpage

\setcounter{equation}{0}
\setcounter{page}{2}

1. Various attempts have been made at combining the ideas of superstring 
unification and the inflationary cosmological scenario. Although the existence
of numerous scalar fields (the moduli) would seem to provide a necessary 
ingredient, the fact that they remain massless at all orders in 
string perturbation theory leads to serious difficulties in  building a 
sucessful cosmological scenario. Among the many moduli, the dilaton ($\phi$) 
is of special interest because it controls the string coupling and variations 
of this field correspond to changes in masses and coupling constants which are
strongly constrained by observation. This problem is usually addressed by 
assuming that the dilaton develops a potential, which must have its origin in
non-perturbative effects, such as  condensation of gauginos, 
so that Einstein's gravity can be recovered after $\phi$ settles into 
the minimum of its potential. However, general arguments show that the dilaton
cannot be stabilized in the perturbative regime of string theory leading
to a runaway problem \cite{dine}.
Moreover, even if the dilaton were stabilized by non-perturbative
potentials, these are too
steep to be suitable for inflation without fine-tuning the initial conditions 
\cite{binetruy}.  Of course, once the dilaton is fixed, inflation can 
be achieved via other fields, e.g. chiral fields (gauge singlets) 
for a suitable choice of 
the inflationary sector of the superpotential \cite{ross, bento}. Furthermore,
there are additional 
difficulties related with the Polonyi problem associated with scalar fields 
that couple only gravitationally and which may dominate the energy 
density of the universe at present \cite{coughlan, banks}.

 In the present letter, we show that the abovementioned difficulties can be 
avoided through a new mechanism introduced in \cite{font1} for fixing the 
dilaton. Indeed, it was shown that the dilaton potential develops a 
suitable minimum once the requirement of S-duality invariance (see below) 
is imposed. Already in Ref. \cite{macorra} this 
mechanism was used to stabilise 
the dilaton while inflation was accomplished by chiral fields. Our analysis 
shows 
that inflation can be achieved via
the dilaton itself provided one considers 
a novel way of implementing the inflationary expansion of the universe 
recently put forward by Linde \cite{linde} and Vilenkin \cite{vilenkin}: 
topological or defect inflation. They have shown that the core of 
a topological defect may undergo exponential inflationary expansion 
provided the scale of symmetry breaking satisfies the condition

\begin{equation}
\label{aa}
\eta > {\cal O} (M_P).
\end{equation}

The inflation that ensues is eternal since the core of the defect is 
topologically stable and it is the restored symmetry in the core that 
provides the vacuum energy for inflation. We shall see that the conditions  
for successful inflation are satisfied by domain walls that separate 
degenerate minima 
in S-dual superstring potentials, implying that topological domain wall
inflation is, as antecipated in Ref. \cite{banks} through general
arguments, an attractive cosmological scenario in string inspired models.

\vskip0.5cm
\noindent
2. S-duality was conjectured \cite{font1} in analogy with a 
well-established symmetry of 
string compactification -- T-duality. Indeed, it was shown that the effective 
supergravity action following from string compactification on orbifolds or 
even Calabi-Yau manifolds is severely constrained by an underlying string 
symmetry, the so-called target space modular invariance 
\cite{ferrara, alvarez}. 
The target space modular group $PSL(2,Z)$ acts on the complex scalar field T 
as 

\begin{equation}
\label{ab}
T \to \frac{a T - i b}{i c T + d}\qquad ;a, b, c, d \in Z, ad-bc=1,
\end{equation}
and $<T>$ is the background modulus associated to the overall scale of 
the internal six-dimensional space on which the string is compactified. 
Specifically, $T=R^2 + i B$, with $R$ being the ``radius'' of the internal 
space and B an internal axion. The target space modular transformation 
contains the well-known duality transformation $R\to 1/R$ as well as discrete 
shifts of the axionic background B and it was shown that this symmetry 
remains unbroken at any order of string perturbation theory \cite{alvarez}. 
The conjectured symmetry would be a further modular invariance symmetry in 
string theory, where the modular group now acts on the complex scalar field 
$S=\phi + i \chi $, where $\chi$ is a pseudoscalar (axion) field. 
This symmetry includes a duality invariance under which the dilaton gets 
inverted (S-duality). S-modular invariance strongly constrains the theory 
since it relates the weak and strong coupling regimes as well as the 
``$\chi$-sectors'' of the theory. 

The form of the N=1 supergravity action including gauge and matter fields 
is specified by the functions $G(\Phi, \Phi^*)=K(\Phi, \Phi^*) + 
\ln \vert W(\Phi)\vert^2$ and $f(\Phi)$; $K(\Phi, \Phi^*)$ is the K\"ahler 
potential and $W(\Phi)$ the superpotential, where  $\Phi$ denotes all chiral 
matter fields. Let us consider, to start with, only two chiral fields $S,
\ T$. At string tree-level, the K\"alher potential for these fields looks like

\begin{equation}
\label{ac}
K = -\ln (S + S^*) - 3 \ln (T + T^*).
\end{equation}

The scalar potential can be written in the following form \cite{font1}

\begin{equation}
\label{ad}
V=\vert h^S \vert^2 G_{S S^*}^{-1} + \vert h^T \vert^2 G_{T T^*}^{-1} - 
3 \exp(G),
\end{equation}
where $h^i=\exp(\frac{1}{2}G) G^i$, $i=S,T$ and the indices denote 
derivatives with respect to the indicated variable.

 We shall first look at the case where there is only one modulus field, $T$. 
It is known that the purely $T$-dependent superpotential 
has to vanish order by 
order in perturbation theory so that the VEV of T remains undetermined. 
However, one expects that non-perturbative effects will generate a 
superpotential for T. The simplest expression for $W(T)$ compatible with 
modular invariance is \cite{ferrara}

\begin{equation}
\label{ae}
W(T) \sim \eta(T)^{-6}
\end{equation}
and the related scalar potential is given by \cite{font2}:

\begin{equation}
\label{af}
V(T) = \frac{1}{T_R^3\vert \eta(T)\vert^{12}}
\left( \frac{T_R^2}{4 \pi^2}\vert \hat G_2(T) \vert^2 -1 \right),
\end{equation}
where $T_R = 2 {\rm Re}\  T$. The function  
$\eta(T)=q^{1/24} \prod_n(1-q^n)$ is the 
well-known Dedekind function, $q\equiv \exp (-2\pi T)$; 
$\hat G_2 = G_2 - 2 \pi/T_R$ is the weight two Eisenstein function 
and $G_2=\frac{1}{3} \pi^2 - 8 \pi^2  \sigma_1(n) \exp(- 2\pi n T)$, where  
$\sigma_1(n)$ is the sum of the divisors of $n$.

If, on the other hand, one considers only the S field, the requirement of 
modular invariance leads to a scalar potential \cite{font1}

\begin{equation}
\label{ag}
V(S) = \frac{1}{S_R\vert \eta(S)\vert^{4}}
\left( \frac{S_R^2}{4 \pi^2}\vert \hat G_2(S) \vert^2 - 3 \right).
\end{equation}

This potential (just like $V(T)$) diverges in the limit $S\to 0, \infty$ and 
has minima at finite values of $S$, close to the critical value $S=1$. Indeed,
the function $\hat G_2(S)$ has its only zeros at $S=1,\ 
S=\exp(\frac{1}{6} i\pi)$
and therefore these self-dual points are necessarily extrema of the potential
(\ref{ag}):

\begin{equation}
\label{aga}
\frac{d V}{d S} = \frac{1}{ \pi S_R\vert \eta(S)\vert^{4}}
\left[ \hat G_2\left( \frac{S_R^2}{4 \pi^2}\vert \hat G_2(S) \vert^2 
     - 3\right) + \frac{S_R}{\pi} \left( \vert \hat G_2(S) \vert^2  + \frac{S_R}{4} (\hat G_{2}^{*} \hat G_{2S} + \hat G_{2S}^{*} \hat G_{2})\right)\right]   = 0;
\end{equation}
more precisely, the former corresponds to a local maximum and the latter to a 
saddle point. The potential has other extrema, namely minima, for 
${\rm Re}\ S\sim 0.8, \ 1.3$ and ${\rm  Im}\ S = n,\ n \in Z$;  these are minima both 
along the ${\rm Re }\ S$ and ${\rm Im}\ S$ directions. The qualitative shape of the 
potential is shown in Fig. 1 and the periodic structure of minima along 
the ${\rm Im}\ S$ direction is shown in Fig. 2.  Therefore, we see that 
the same way target space modular 
invariance fixes the value of $T_R$ thus forcing the theory to be 
compactified, S-modular invariance fixes the value of $S_R$ thus 
stabilizing the potential and avoiding the dilaton runaway problem. It is 
clear that  the theory has to choose among an infinity of degenerate 
minima whose 
positions differ by modular transformations. Once one of them is chosen, 
target space modular invariance is spontaneously broken. Since duality 
is a discrete symmetry, if there were a phase in the evolution of the universe 
in which the compactification radius was already
spontaneously chosen, S-duality domain 
walls would be created separating different vacua.

 A  more realistic model is obtained once all gauge 
singlet fields of the theory: $S, T_i$, $i=1, 2, 3$ are considered. 
Imposing S and T-duality on the 
Lagrangian for $N=1$ supergravity theory, one  obtains the following
potential \cite{macorra}:

\begin{equation}
\label{bca}
 V=e^K \vert\eta(T_2)\eta(T_3)\eta(S)\vert^{-4}
\left(\vert P\vert^2\left[\frac{S_R^2}{4\pi^2}\vert\hat G_2(S)\vert^2 + 
\frac{T_{Ri}^2}{4\pi^2}\vert\hat G_2(T_i)\vert^2 - 2\right] + F_0\right),
\end{equation}
where $P=P(T_1,\psi)=\lambda(T_1)\Theta(\psi)$, $\psi$ denotes the 
untwisted chiral fields related to the $T_1$ sector, $\lambda$ and $\Theta$ 
are gauge invariant functions and $F_0=P_m(K^{-1})_n^mP^n +
(K_mP(K^{-1})_{T_1}^mP^{T_1} + h.c.)$.

Clearly, this potential is S (and T)-duality invariant since all dependence 
on $S$ is given in terms of duality-invariant functions 
$e^K \vert\eta(S)\vert^{-4}$ and $S_R^2\vert \hat G_2(S)\vert^2$. Again, the 
dual invariant points $<S>=1,\ e^{-\pi/6}$ and 
$<T_i>=1,\ e^{-\pi/6}$ are extrema (maxima and saddle points, respectively) 
and the minima of V are nearby.
In what follows we will show that the 
conditions for topological inflation to occur at the core of 
the domain walls separating degenerate minima of the above 
potential can be met for some range of parameters.

\vskip0.5cm
\noindent
3. Let us now turn to the discussion of the conditions 
for successful topological inflation. Along a domain wall 
$\chi$ ranges from one minimum in one region 
of space to another minimum in another region. Somewhere between, $\chi$ 
must traverse the top of  the potential, $\chi\approx \chi_0$ and
we hence start expanding the potential about $\chi_0$

\begin{equation}
\label{ah}
V\approx V_0 \left( 1-\alpha^2 \frac{(\chi - \chi_0)^2}{M^2}\right),
\end{equation}
where $M=M_P/\sqrt{8\pi}$, which is the natural scale of the fields in supergravity
and was set to one in the previous section.

In flat space, the wall thickness is equal to the curvature of the 
effective potential, that is $\delta^{-1} \sim \alpha(V_0/M_P^2)^{1/2}$. The 
Hubble parameter in the interior of the wall is given by
$H\approx (8\pi G V_0/3)^{1/2}$. If 
$\delta \ll H^{-1}$, gravitational effects are negligible. However, if 
$\delta > H^{-1}$, the region of false vacuum near the top of the potential, 
$V\approx V_0$, extends over a region greater than a Hubble volume. Hence, if 
the top of the potential satisfies the conditions for inflation, the interior 
of the wall inflates. Demanding that $\delta > H^{-1}$, one obtains the 
following condition on $\alpha$, 

\begin{equation}
\label{ai}
\alpha^2 < 8 \pi/3 .
\end{equation}

It turns out that this condition  is more stringent than the ones 
that can be derived from demanding an inflationary slow rollover regime
\cite{banks}. However, the requirement that there are at least $N_e$ e-folds 
of inflation, i.e

\begin{equation}
\label{ba}
\frac{-V^{\prime\prime}}{V}\ll \frac{6 \pi}{N_e}
\end{equation}
leads to the most stringent constraint on $\alpha$ (we assume here $N_e=65$)
\cite{banks}

\begin{equation}
\label{bb}
\alpha^2 < 3\pi /65.
\end{equation}

We have computed $\alpha^2$ for the purely T-dual and S-dual potentials, Eqs.
(\ref{af}) and (\ref{ag}), assuming that the real part of the (T,\ S) field 
has already settled at the minimum of the potential. We envisage a scenario 
where inflation would take place at the core of domain walls separating 
different vacua, as the imaginary part of the (T,\ S) field expands 
exponentially once the conditions discussed above are satisfied at the top 
of potential (see Fig.\ 2). We  find $\alpha^2=0.30$ and $\alpha^2=0.09$, 
respectively; hence, we conclude that the conditions for successful defect 
inflation to occur are fulfilled in the purely S-dual case only.
 
For the more realistic model of Eq. (\ref{bca}), the value of $\alpha^2$ 
depends on the vacuum contributions of $\vert P\vert^2$ and $F_0$. 
In this case, we assume that  the 
T-fields and the untwisted fields of the $T_1$ sector have already settled 
at the minimum of the potential and inflation takes place due to 
the S-field; the relevant potential can be then written as 

\begin{equation}
\label{be}
V(S) =\frac{1}{S_R\vert \eta(S)\vert^{4}}
\left( \frac{S_R^2}{4 \pi^2}\vert \hat G_2(S) \vert^2 - a \right),
\end{equation}
where $a$ is a constant. As for the models discussed above, we shall 
further assume that $S_R$ has settled at the minimum of the potential (at
$<S_R>= 2.$) and inflation would take place at the core of domain walls 
that separate different vacua, along the ${\rm Im}\ S$ direction. 
We have computed $\alpha^2$  for 
different values of $a$ and found that the condition (\ref{bb}) is not 
fulfilled if $F_0 \ge 0$. However, if the vacuum contribution of untwisted 
chiral fields is such that $F_0 < 0$, actually when  $a \gaq\ 2.5 $, 
then successful topological inflation
can occur in realistic models as well. Notice that we have included in our computation the effect of  the non-canonical struture of the kinetic terms
of $S$ (and $T$)  dictated by 
N=1 supergravity , $(S_R)^{-2} \partial_\mu S\partial^\mu S^{*}$ 
$((T_R)^{-2} \partial_\mu T\partial^\mu T^{*})$.
Hence, we conclude that topological inflation is possible for $a \gaq\ 2.5 $, 
thereby solving the initial condition problems in these models.

Of course, in order to have a complete cosmological scenario, 
it is still required that primordial energy
density fluctuations are generated and a successful phase of reheating is 
achieved. For that, we notice that the $S$ field is coupled to radiation
through Re $\ f(\Phi)$, i.e. via the term ${\rm Re} S
F_{\mu\nu}^{a}F^{\mu\nu a}$. Let us now see how  
domain wall evolution and dynamics  leads, as discussed in Refs. 
\cite{linde, vilenkin, linlin}, to a consistent cosmological scenario.
At the top of the potential the field S will be submitted to large fluctuations
within time $\Delta t=H^{-1}$. These fluctuations look like sinusoidal waves 
with amplitude $H/2\pi$ and wavelength $O(H^{-1})$. During this time
the original domains inflate about $e$ times and therefore the horizon,
whose dimension is about $H^{-1}$, 
becomes divided into about $e^3$ domains. The presence of horizons for de
Sitter spaces implies that the evolution of the field in each domain is 
independent of the others. In half of the domains the average values of the 
field is given by $<\chi>_T = H/2\pi$ and $\chi$ moves towards 
$\bar \chi \equiv <{\rm Im}\ S> = n$,
the VEV of ${\rm  Im}\ S$ (cf. Figure 1), while for
the other half  $<\chi>_T=-H/2\pi$ and $\chi$ moves towards $\bar \chi=n + 1$
or $\bar \chi=n - 1$. 
Since for small values of the field its dynamics is dominated 
by quantum fluctuations, the 
universe becomes divided into many thermalized regions surrounded by 
exponentially inflating domain walls. That is, the regime dominated
by quantum fluctuations allows, as in the new inflationary model, that
primordial energy density fluctuations are generated and reheating to occur.
The domain walls create, on their turn, new
inflating walls as,  due to fluctuations, domains where for instance  
$\bar \chi=n + 1$ or $\bar \chi=n - 1$ can
be produced inside a domain where $\bar \chi =n$. The latter regions 
will then be like
islands of  $\bar \chi=n + 1$ or $\bar \chi=n - 1$
in a sea of $\bar \chi=n$. Thus, the 
evolution of the walls does not destroy the original domain walls but, on the
contrary,  creates new domain walls on a smaller scale. These new walls will
be generated in regions where $ \chi \approx \chi_0$ such 
that a jump of $\chi$ from one vacuum to another
is not too unlikely. This implies that the new domain 
walls will be predominantly created close to the older ones, leading in this
way to a self-similar fractal domain wall structure \cite{vilenkin, linlin}.  
Due to the domain wall classical
dynamics, once topological inflation starts it never ends and that ensures that
the S field is free of any postmodern Polonyi type problem.  

\vskip0.5cm
\noindent
4. In conclusion, we have shown that topological inflation is a viable 
alternative to achieve inflation in string models with S-duality. Our study 
shows that in purely S-dual models the potential does allow topological 
inflation to occur due to the domain
walls separating the different degenerate $\chi
$-sectors of the theory. This
does not happen for purely T-dual models. For models
where both S and T duality are imposed, defect inflation triggered by domain
walls occurs only for a limited
range of the parameters associated with the vacuum contributions of untwisted 
fields to the potential, i.e.  for $F_0<0$ (cf. Eq. (\ref{bca})) and 
$a \gaq\ 2.5 $ in the model of Eq. (\ref{be}).


\newpage

{\bf Figure Captions}

\vskip 2cm
{\bf Fig. 1} The scalar potential V for the purely S-dual model, Eq. 
(\ref{ag}), as a function of (Re S, Im S).

\vskip 2cm
{\bf Fig.2} The periodic structure of minima of V for the purely S-dual model 
along the Im S direction, for  Re S at its  minimum $(\sim 1.3)$.


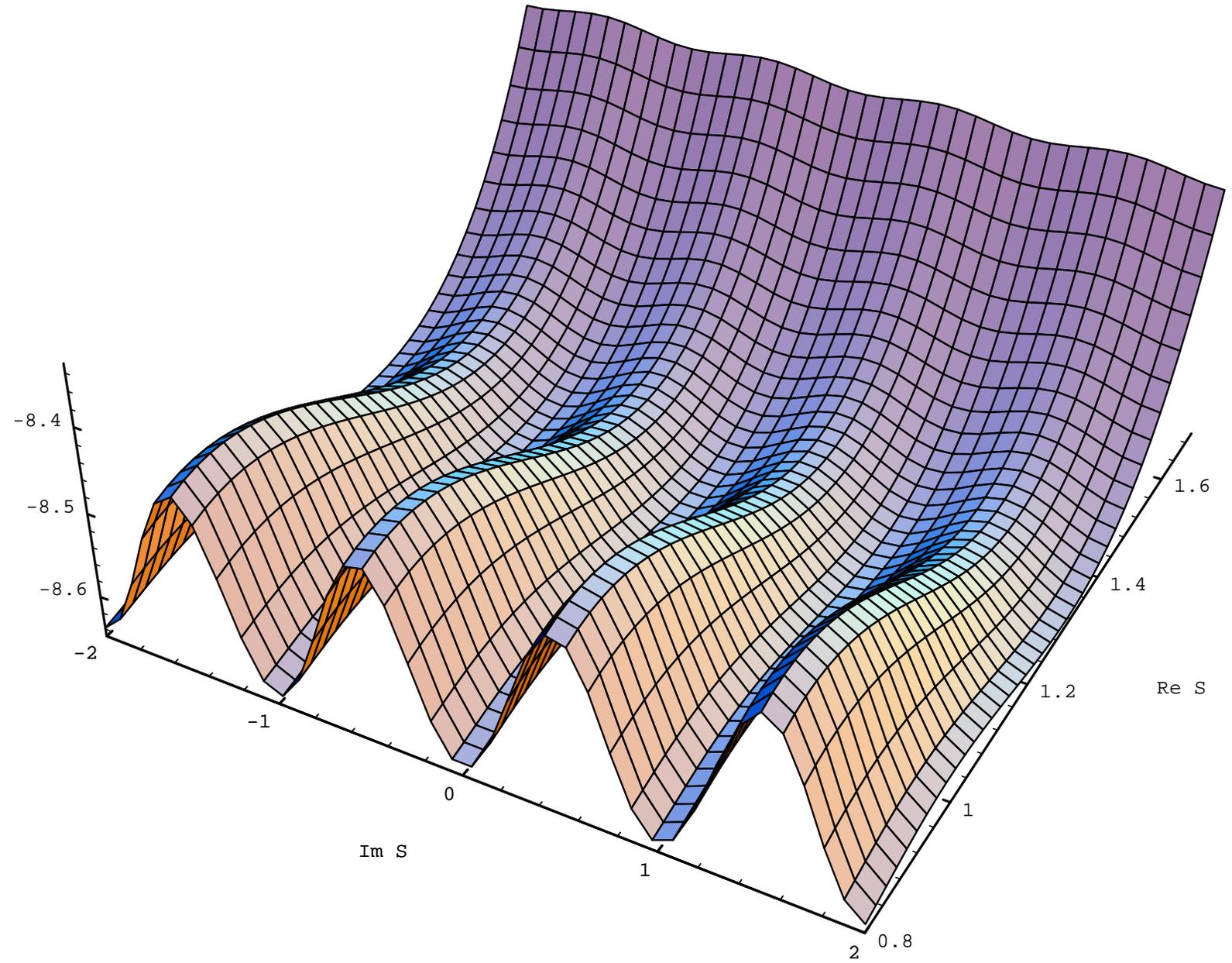
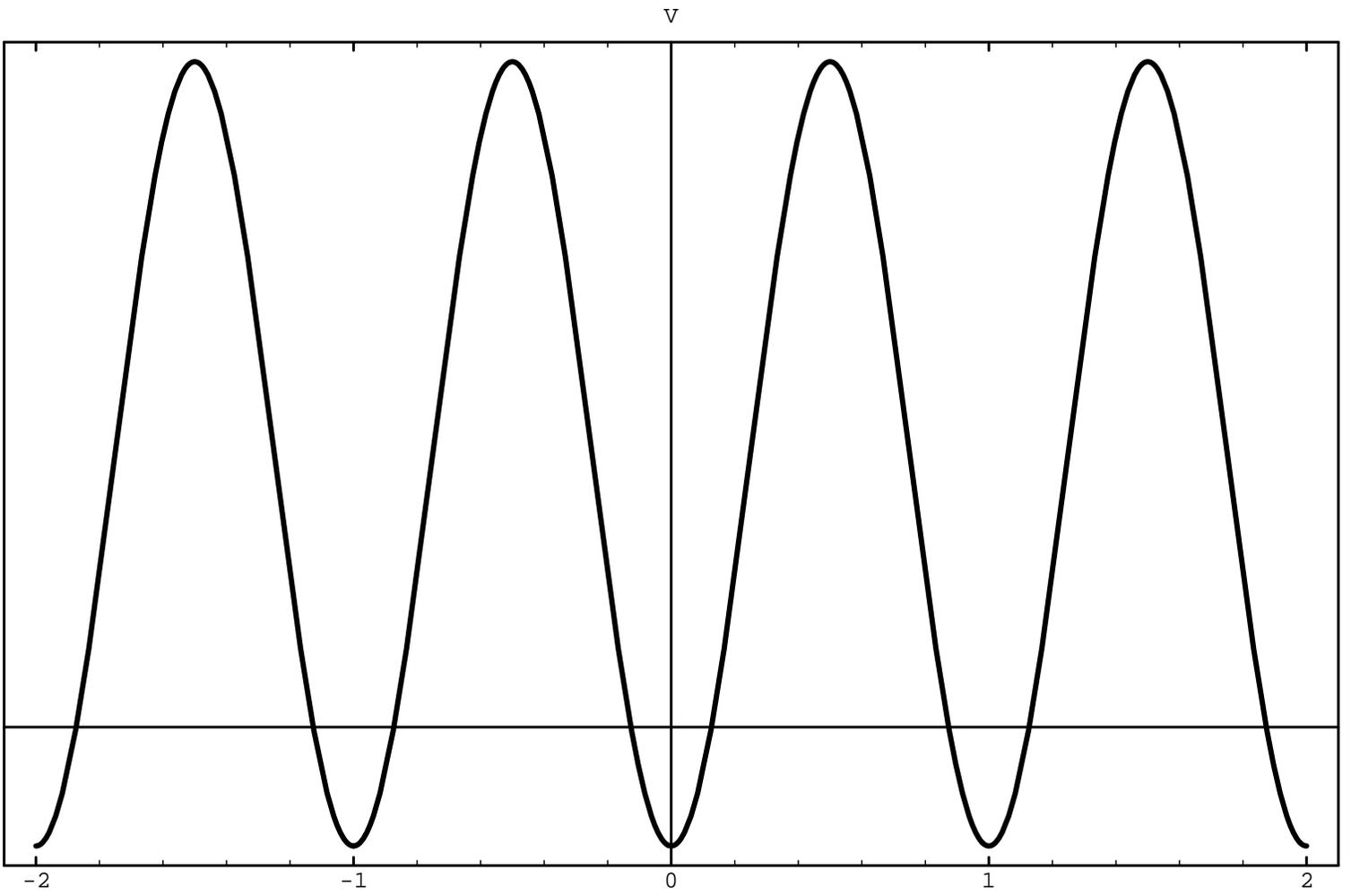
\end{document}